\documentclass{article}

\usepackage{epsfig}
\usepackage{subfigure}
\usepackage{calc}
\usepackage{graphicx}

\usepackage{amsfonts}
\usepackage{amsmath}
\usepackage{amssymb}
\usepackage{latexsym}

\begin{document}

\title{Correction to ``Generalized Self-Shrinking Generator''}
\date{}
\author{Amparo F\'{u}ster-Sabater\\
{\small Instituto de F\'{\i}sica Aplicada, C.S.I.C.}\\
{\small Serrano 144, 28006 Madrid, Spain} \\
{\small amparo@iec.csic.es}}

\maketitle

\begin{abstract}

In this correspondence, it is given a correction to
\textit{Theorem 4} in Y. Hu, and G. Xiao, ``Generalized
Self-Shrinking Generator,'' \textit{IEEE Transactions on
Information Theory}, vol. 50, No. 4, pp. 714-719, April 2004.


\end{abstract}

\section{Introduction}
\footnotetext{This work was supported in part by CDTI (Spain)
 and the companies INDRA, Uni\'{o}n Fenosa, Tecnobit, Visual Tools, Brainstorm, SAC and Technosafe under Project Cenit-HESPERIA
 as well as supported by Ministry of Science and Innovation and European FEDER Fund under Project TIN2008-02236/TSI.} The purpose
 of this note is to point out
that \textit{Theorem 4} in \cite{Hu} is not valid in all cases.
The statement of the theorem reads:

[1, Theorem 4]: \textit{No more than $1/4$ of the sequences from
$B(a)$ have least periods less than $2^{n-1}$}.

In the following, two counter examples of \textit{Theorem 4} and a
reformulation of such a theorem are given.

\textit{Counter example 1}: We take the $n = 3$ degree
\textit{m}-sequence

\begin{center}
$a$ = 1110010 $\sim$
\end{center}
whose minimal polynomial is $x^3 + x^2 + 1$. Then we get $B(a)$
i.e., the family of $2^3$ generalized self-shrinking sequences
based on $a$ (see \cite{Hu}):
\begin{center}
 1.  $G$ = (000), $\{b(G)\}$ = 0000 $\sim$ \\
 2.  $G$ = (100), $\{b(G)\}$ = 1111 $\sim$ \\
 3.  $G$ = (010), $\{b(G)\}$ = 0110 $\sim$ \\
 4.  $G$ = (110), $\{b(G)\}$ = 1001 $\sim$ \\
 5.  $G$ = (001), $\{b(G)\}$ = 1010 $\sim$ \\
 6.  $G$ = (011), $\{b(G)\}$ = 1100 $\sim$ \\
 7.  $G$ = (101), $\{b(G)\}$ = 0101 $\sim$ \\
 8.  $G$ = (111), $\{b(G)\}$ = 0011 $\sim$ \\
\end{center}

In $B(a)$, there are 4 generalized self-shrinking sequences with
least period $T = 2^2$, i.e., $\{$0110 $\sim$, 1001 $\sim$, 1100
$\sim$, 0011 $\sim$$\}$. Nevertheless, there are $|B'|$ = 4
generalized self-shrinking sequences with periods less than $2^2$,
i.e., $\{$0000 $\sim$, 1111 $\sim$$\}$ with least period $T = 1$
and $\{$0101 $\sim$, 1010 $\sim$$\}$ with least period $T = 2$.
Thus, $1/2$ of the sequences from $B(a)$ have least period less
than $2^2$, contradicting the claimed result. Such a contradiction
can be justified as follows:

In the proof of \textit{Theorem 4} in \cite{Hu}, it is stated that
$b(v^{(1)})+ B'$, $b(v^{(2)})+B'$ and $b(v^{(3)})+B'$ are
different cosets of $B'$.

Nevertheless, in this particular example we have:

$B' = \{$0000 $\sim$, 1111 $\sim$, 0101 $\sim$, 1010 $\sim\}$

$b(v^{(1)}) = b($010$) = \{$0110 $\sim\}$

$b(v^{(2)}) = b($011$) = \{$1100 $\sim\}$

$b(v^{(3)}) = b($001$) = \{$1010 $\sim\}. \\$

Therefore,

$b(v^{(1)})+ B' = \{$0110 $\sim$, 1001 $\sim$, 0011 $\sim$, 1100
$\sim \}$

$b(v^{(2)})+ B' = \{$1100 $\sim$, 0011 $\sim$, 1001 $\sim$, 0110
$\sim \}$

$b(v^{(3)})+ B' = \{$1010 $\sim$, 0101 $\sim$, 1111 $\sim$, 0000
$\sim\}. \\$

Thus, the set of generalized self-shrinking sequences $b(v^{(1)})+
B'$ equals $b(v^{(2)})+ B'$ as well as $b(v^{(3)})+ B'$ equals
$B'$. Analogous results can be obtained for the reverse version of
the 3 degree \textit{m}-sequence $a$. Consequently, for $n = 3$
there are no three different sets of generalized self-shrinking
sequences with least period $T = 2^{n-1}. \\$

\textit{Counter example 2}: We take the $n = 2$ degree
\textit{m}-sequence

\begin{center}
$a$ = 110 $\sim$
\end{center}
whose minimal polynomial is  $x^2 + x + 1$. Then we get $B(a)$
i.e., the family of $2^2$ generalized self-shrinking sequences
based on $a$ (see \cite{Hu}):

\begin{center}
 1.  $G$ = (00), $\{b(G)\}$ = 00 $\sim$ \\
 2.  $G$ = (10), $\{b(G)\}$ = 11 $\sim$ \\
 3.  $G$ = (01), $\{b(G)\}$ = 01 $\sim$ \\
 4.  $G$ = (11), $\{b(G)\}$ = 10 $\sim$ \\
\end{center}

In $B(a)$, there are now 2 generalized self-shrinking sequences
with least period $T = 2$, i.e., $\{$01 $\sim$, 10 $\sim \}$.
Nevertheless, there are $|B'| = 2$ generalized self-shrinking
sequences with period less than 2, i.e., $\{$00 $\sim$, 11 $\sim
\}$ with least period $T = 1$. Thus, $1/2$ of the sequences from
$B(a)$ have least periods less than 2, contradicting the claimed
result. Such a contradiction can be justified as before:

In the proof of \textit{Theorem 4} in \cite{Hu}, it is stated that
$b(v^{(1)})+ B'$, $b(v^{(2)})+ B'$ and $b(v^{(3)})+ B'$ are
different cosets of $B'$. In this particular example:$\\$

$B' = \{$00 $\sim$, 11 $\sim\}$

$b(v^{(1)}) = b($01$) = \{$01 $\sim\}$

$b(v^{(2)}) = b($11$) = \{$10 $\sim\}$

$b(v^{(3)}) = b($10$) = \{$11 $\sim\}. \\$

Therefore,

$b(v^{(1)})+ B' = \{$01 $\sim$, 10 $\sim \}$

$b(v^{(2)})+ B' = \{$10 $\sim$, 01 $\sim \}$

$b(v^{(3)})+ B' = \{$11 $\sim$, 00 $\sim \}. \\$

Thus, the set of generalized self-shrinking sequences $b(v^{(1)})+
B'$ equals $b(v^{(2)})+ B'$ as well as $b(v^{(3)})+ B'$ equals
$B'$. Consequently, for $n = 2$ there are no three different sets
of generalized self-shrinking sequences with least period $T
=2^{n-1}$.

For $n \geq 4$, the theorem holds (see the example given in
\cite{Hu} for $n = 4$) as the number of sequences from $B(a)$ with
least periods $T < 2^{n-1}$ is at most $1/4$ of $|B(a)|$.
Therefore, the sets of generalized self-shrinking sequences
$b(v^{(1)})+ B'$, $b(v^{(2)})+ B'$ and $b(v^{(3)})+ B'$ are
actually different cosets of $B'$.

In brief, the result in \textit{[1, Theorem 4]} regarding the
number of generalized self-shrinking sequences having least period
less than $2^{n-1}$ is not affected for values of $n \geq 4$. In
this way, \textit{Theorem 4} can be reformulated as follows:$\\$

[1, Theorem 4 - Revised]: \textit{If $n \geq 4$, then no more than
$1/4$ of the sequences from $B(a)$ have least periods less than
$2^{n-1}$}.

\end{document}